# Numerical exploration on unveiling the photovoltaic potential of $MgXS_3$(X = Ti, Zr, Hf) chalcogenide perovskites

Md. Mizanur Rahman[1,2], Md. Nahid Hasan[2], Tanvir Ahmed[2,3], and Jaker Hossain[2*]

[1]*Institute of Environmental Science, University of Rajshahi, Rajshahi 6205, Bangladesh.*

[2]*Photonics & Advanced Materials Laboratory, Department of Electrical and Electronic Engineering, University of Rajshahi, Rajshahi 6205, Bangladesh.*

[3]*Department of Electrical & Electronic Engineering, First Capital University of Bangladesh, Chuadanga 7200, Bangladesh.*

**Abstract**

Lead-free chalcogenide perovskites offer a nontoxic and thermally robust path beyond Pb-based perovskite solar cells (PSCs), but their device-level behavior in realistic three-dimensional geometries remains insufficiently characterized. In this work, we investigate ZnSe/$MgXS_3$(X = Ti, Zr, Hf)/$Sb_2S_3$ solar cell architecture where $MgXS_3$ absorbers from the II–IV–VI chalcogenide perovskite family is employed as the absorber layer. The device is analyzed using 3D finite-element simulations in COMSOL Multiphysics that self-consistently couple optical generation, drift–diffusion carrier transport, and heat transfer under AM 1.5G 1-sun illumination, following a fully coupled opto–electro–thermal framework. For each absorber composition, the impacts of absorber thickness, doping, and defect density are systematically investigated, and the contribution of an $Sb_2S_3$ back-surface-field (BSF) layer to carrier collection and spectral response is quantified. Under optimized conditions, $MgZrS_3$, $MgTiS_3$, $MgHfS_3$-based devices achieves a simulated power conversion efficiency (PCE) of 28.18%, 26.72%, and 28.16%, respectively. The corresponding open-circuit voltage ($V_{OC}$) values are 0.94 V, 0.74 V, and, 1.07 V while the short-circuit current density ($J_{SC}$) values are 34.46 mA/cm$^2$, 42.69 mA/cm$^2$, and 29.89 mA/cm$^2$, with fill factor (FF) values of 86.99%, 84.58%, and 88.02%, respectively. Coupled electro-thermal simulations further reveal a small spatially non-uniform steady-state temperature rise across the ultrathin cell stack, mainly governed by non-radiative recombination and Joule dissipation within the active layers. Overall, these results confirm $MgXS_3$(X = Ti, Zr, Hf) chalcogenide perovskites as promising lead-free absorber materials and offer practical design guidance for achieving high-efficiency, thermally stable three-dimensional device architectures.

## 1. Introduction

The growing demand of environment friendly, sustainable as well as efficient energy solutions has significantly accelerated research into advanced photonic devices. Among these, perovskite solar cells (PSCs) are widely recognized as strong candidates for next-generation photovoltaics, offering notably power conversion efficiencies, low manufacturing cost, long charge-carrier diffusion lengths, low exciton binding energies, and composition-tunable bandgaps. In addition, solution-based processing and high optical absorption make them attractive for scalable device design [1-3].

Maximizing photovoltaic efficiency while minimizing energy losses remains a central challenge in solar-energy materials. Organic–inorganic lead-halide perovskite solar cells (LHPSCs) emerged as a third-generation platform addressing cost and performance limits of earlier technologies and rapidly achieved a certified efficiency of 26.1% [4], lead-based perovskites with compositions such as $CH_3NH_3PbI_3$ and $NH_2CHNH_2PbI_3$ demonstrate notable performance [5-6]. However, large-scale deployment remains challenging because lead toxicity raises environmental and health concerns, including the risk of soil and water contaminations. In addition, limited stability under prolonged exposure to moisture, heat, and UV radiation reduces long-term durability in outdoor operation and remains a key barrier to commercialization [7-9], and mechanistic studies implicates the organic (A-site) component in decomposition pathways [10]. These limitations have driven interest in lead-free, fully inorganic perovskites, particularly chalcogenide perovskites, as promising candidates that may offer improved stability while maintaining competitive photovoltaic performance.

Chalcogenide perovskites (CPs) have emerged as promising, lead-free and environmentally benign photovoltaic materials, motivated by the instability and toxicity concerns of hybrid halide perovskites. Combining the favorable electronic properties of halides with the chemical and thermal robustness of oxides, CPs exhibit strong optical absorption and suitable semiconducting behavior for use as solar absorbers [11-14]. In particular, CPs which follow the general formula $ABX_3$ where A is a large cation, such as $Ca^{2+}$, $Mg^{2+}$, $Ba^{2+}$, $Zn^{2+}$ etc., B is smaller transition metal cation namely, $Sn^{4+}$, $Hf^{4+}$, $Zr^{4+}$, $Ti^{4+}$ and X is chalcogen anion ($S^{2-}$ or $Se^{2-}$) have reaped attention due to their stability, non-toxicity and availability on earth [15-16]. Moreover, across a range of CPs, spectroscopic-limited maximum efficiency (SLME) analyses together with band-structure and absorption data indicate photovoltaic-appropriate bandgaps and high absorption coefficients, confirming their suitability as efficient solar absorbers [17].

Recent first-principles studies have identified the $MgXS_3$ (X = Ti, Zr, Hf) chalcogenide perovskites as promising lead-free absorber candidates for single-junction photovoltaics, within the broader class of $AB(S,Se)_3$ perovskites proposed as stable alternatives to Pb-halide perovskites [18-20]. Structurally, $MgTiS_3$, $MgZrS_3$, and $MgHfS_3$ crystallize in the $CaZrS_3$-type perovskite framework with an orthorhombic Pnma phase, built from corner-sharing $XS_6$ octahedra and Mg occupying the A-site positions. The calculated lattice constants typically fall in the ranges a ≈ 6.6–6.7 Å, b ≈ 7.0 Å, and c ≈ 9.2–9.6 Å. Their bulk moduli, in the order of 80–250 GPa, suggest mechanically robust lattices that are well suited for thin-film device fabrications [18-19]. Goldschmidt tolerance factors of roughly 0.75–0.79 place these compounds within the established stability window for distorted perovskites, further supporting their structural viability as photovoltaic absorbers [19,21]. Electronically, the $MgXS_3$ (X = Ti, Zr, Hf) compounds are semiconductors with bandgaps well suited for single-junction photovoltaics. In particular, $MgTiS_3$ and $MgZrS_3$ exhibit near-direct or direct transitions with gaps of approximately 1.10 eV and 1.30 eV at the Brillouin-zone center [18], while $MgHfS_3$ lies near 1.43 eV, close to the detailed-balance optimum for high-efficiency devices [19,22]. Density-of-states analyses reveal strong hybridization between Mg-2p, X-d (Ti/Zr/Hf) and S-2p orbitals near the band edges, yielding reasonably dispersive bands and suggesting adequate carrier transport for device operation [18-19,22]. Optically, $MgXS_3$ chalcogenides exhibit relatively high static dielectric permittivities, with $\varepsilon_0$ of approximately 17 for $MgTiS_3$, 32 for $MgZrS_3$, and 9.6 for $MgHfS_3$, together with strong near-visible absorption. Reported absorption coefficients reach the $10^6$ $cm^{-1}$ range, with representative values of about $2.73 \times 10^6$ $cm^{-1}$ for $MgTiS_3$, $5.02 \times 10^6$ $cm^{-1}$ for $MgZrS_3$, and $5.12 \times 10^6$ $cm^{-1}$ for $MgHfS_3$ [18,22]. These data imply that $MgXS_3$ films only a few hundred nanometers thick can harvest most above-band-gap photons. Within this family, the larger absorption coefficients of $MgZrS_3$ and $MgHfS_3$ indicate that, for a given thickness, they will absorb more strongly than $MgTiS_3$ and therefore are particularly attractive for ultra-thin device architectures [18,22]. Spectroscopic-limited maximum efficiency analyses based on the calculated optical spectra predict theoretical efficiencies exceeding ~29–32% for optimized $MgXS_3$ absorbers [18], underscoring the potential of this lead-free chalcogenide perovskite family as next-generation photovoltaic materials.

Perovskite solar cells are typically arranged as a selective-contact stack with a window at the illuminated side and a back-surface field at the rear. In this design, ZnSe serves as the window. Its wide bandgap of about 2.7 eV, electron mobility of approximately 50 $cm^2 V^{-1} s^{-1}$, and electron affinity near 4.09 eV yield a small, extraction-friendly conduction-band offset with the $MgXS_3$ (X = Ti, Zr, Hf) absorbers whose electron affinity is about 4.1 eV [23]. ZnSe also offers a high refractive index and a polycrystalline microstructure that together support strong transmission across the visible and near-ultraviolet spectrum, and its low reactivity toward humidity is advantageous for manufacturing [23]. In contrast, a back surface field at the rear contact establishes an electric field that enhances minority carrier reflection, strengthens carrier collection, and suppresses recombination. $Sb_2S_3$ is well suited for this role. Its bandgap of about 1.62 eV and electron affinity near 3.7 eV yield a favorable valence-band alignment that blocks electrons and promotes hole extraction, while a hole mobility on the order of 10 $cm^2 V^{-1} s^{-1}$ supports low-loss transport. $Sb_2S_3$ also offers chemical stability and is composed of earth-abundant antimony and sulfur, making it an attractive, sustainable choice for a back surface field in photovoltaic devices [24].

While first-principles studies have successfully predicted the exceptional intrinsic optoelectronic properties of $MgXS_3$ chalcogenide perovskites [18-19,22], translating these macroscopic material parameters into viable device-level efficiencies require accounting for real-world geometric and spatial bottlenecks. Numerous free photovoltaic modeling tools are available, but most are restricted to one-dimensional device stacks and lack Multiphysics coupling, time dependence, and robust post-processing, so they cannot capture a solar cell's three-dimensional features [25-27]. Popular 1D platforms such as SCAPS and AMPS therefore miss lateral nonuniformities and, importantly, do not support thermal analysis that maps heat generation, dissipation to the environment, or conduction across adjacent layers. Furthermore, traditional one-dimensional numerical solvers assume perfectly uniform, planar configurations, thereby neglecting lateral carrier-diffusion gradients, localized and pinhole defects [28-29]. To bridge this gap, this work introduces a methodological advance by developing a fully coupled 3D finite-element simulation framework in COMSOL Multiphysics for $MgXS_3$ perovskite solar cells. The FEM-based approach excels flexible unstructured meshing and accurate modeling of multilayer thin-film PV structures, including absorber, transport-layer, and interface regions. Moreover, adaptive mesh refinement can be applied locally at critical interfaces, such as the ETL/absorber and absorber/HTL junctions, where steep gradients in carrier concentration and electric field occur, thereby enhancing numerical accuracy without significantly increasing computational cost [30-31]. COMSOL Multiphysics provides a Semiconductor Module for drift–diffusion transport, an Electromagnetic Waves interface for optical field and photogeneration calculations, and a Heat Transfer Module for temperature mapping. These coupled physics enable a self-consistent 3D description of photogeneration, recombination, carrier transport, electrostatic potential, and heat flow under operating conditions [32]. In this work, we employ a 3D COMSOL framework to represent real device geometry and material parameters, apply mesh strategies tailored to each sub-layer, and incorporate contact properties and defect densities to reflect realistic operating environments for $MgXS_3$ perovskite solar cells. This approach yields geometry-aware predictions of performance and physical insight that are inaccessible to 1D tools, and it supports the rational development of non-toxic, stable, and power-efficient $MgXS_3$ photovoltaic devices.

## 2. Device layout and modeling strategy

### 2.1 Structure of the proposed photovoltaic device

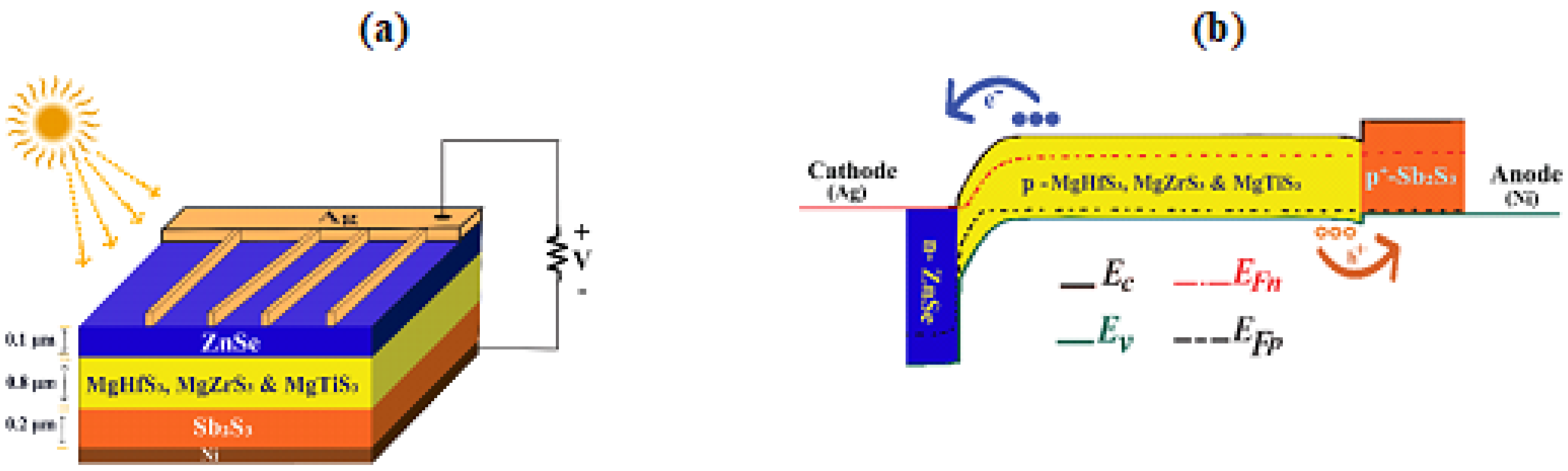

**Fig. 1.** (a) Schematic device configuration and (b) illuminated energy band diagram of the n-ZnSe/p-$MgXS_3$(X = Ti, Zr, Hf)/$p^+$- $Sb_2S_3$ heterojunction perovskite solar cells.

Fig. 1(a) shows the device architectures examined in this work. Each stack uses ZnSe as the window layer and $Sb_2S_3$ as the back surface field layer. The absorber layer is varied among three candidates: $MgHfS_3$, $MgZrS_3$, and $MgTiS_3$. Fig. 1(b) represents the energy band diagram for the sequence Ag/ZnSe/$MgXS_3$(X=Ti, Zr, Hf)/$Sb_2S_3$/Ni. To ensure a consistent evaluation, the same device architecture have been used for all three absorbers. The conduction band offset (CBO) at the ZnSe/$MgXS_3$ interface and the valence band offset (VBO) at the $MgXS_3$/$Sb_2S_3$ interface are calculated from the electron affinities and bandgaps of the constituent materials [33]. The ZnSe/$MgXS_3$ junction exhibits a near-flat conduction band alignment with a CBO of approximately 0.01 eV, facilitating efficient electron extraction without introducing a significant transport barrier. Similarly, the $MgXS_3$/$Sb_2S_3$ interface yields VBO values of 0.12 eV, −0.08 eV, and −0.21 eV for $MgTiS_3$, $MgZrS_3$, and $MgHfS_3$, respectively. Since to get maximum performance, the magnitudes of CBO or VBO should remain within the generally accepted range ($|\Delta E| < 0.3$ eV), severe cliff- or spike-type band discontinuities are avoided, thereby promoting efficient carrier collection and minimizing interfacial recombinations [34]. These results provide physical justification for the selection of ZnSe and $Sb_2S_3$ as the window and BSF layers in the proposed device configuration.

The junction between the window and the absorber forms the primary p-n interface that drives the separation and collection of photocarriers. The interface between the absorber and the back surface establishes a rear electric field that reflects minority carriers into the absorber and reduces recombination at the rear surface. Silver (Ag) serves as the transparent front contact and acts as the cathode. Nickel (Ni) serves as the back contact and acts as the anode. These metals were selected to support the intended band alignment with their adjacent layers and to enable ohmic or near-ohmic behavior under the simulated conditions.

The photovoltaic devices were simulated using COMSOL Multiphysics, and the input material parameters are scratched from published literature [18, 22-24, 35]. The thermophysical parameters used in the Heat Transfer in Solids module, namely thermal conductivity, specific heat capacity, and density, were taken from available literature reports for ZnSe and $Sb_2S_3$, while those for $MgXS_3$ (X = Ti, Zr, Hf) were adopted from reported first-principles/thermodynamic data and reasonable estimates based on closely related chalcogenide perovskites where direct experimental values are not available [19, 36-40]. The essential simulation parameters are concise in Table 1.

**Table 1:** Input device parameters used in the 3D simulation.

| Parameters | ZnSe | $MgTiS_3$ | $MgZrS_3$ | $MgHfS_3$ | $Sb_2S_3$ |
|---|---|---|---|---|---|
| Depth (μm) | 0.1 | 0.8 | 0.8 | 0.8 | 0.2 |
| Energy gap, $E_g$ (eV) | 2.7 | 1.1 | 1.3 | 1.43 | 1.62 |
| Affinity of Electron, χ (eV) | 4.09 | 4.1 | 4.1 | 4.1 | 3.7 |
| Dielectric Permittivity (relative) | 10 | 17 | 32 | 9.6 | 7.08 |
| CB effective DOS ($cm^{-3}$) | $1.8\times 10^{19}$ | $1.8\times 10^{18}$ | $7.83\times 10^{18}$ | $1.8\times 10^{19}$ | $1\times 10^{19}$ |
| VB effective DOS ($cm^{-3}$) | $1.5\times 10^{18}$ | $2.2\times 10^{18}$ | $4.76\times 10^{18}$ | $2.2\times 10^{18}$ | $2\times 10^{19}$ |
| Electron mobility ($cm^2$/(V.s) | 50 | 10 | 10 | 10 | 9.8 |
| Hole mobility ($cm^2$/(V.s) | 20 | 5 | 5 | 5 | 10 |
| Donor concentration, $N_D$ ($cm^{-3}$) | $10^{18}$ | 0 | 0 | 0 | 0 |
| Acceptor concentration, $N_A$ ($cm^{-3}$) | 0 | $10^{18}$ | $10^{18}$ | $10^{18}$ | $10^{19}$ |
| Defect concentration ($cm^{-3}$) | $10^{14}$ | $10^{14}$ | $10^{14}$ | $10^{14}$ | $10^{14}$ |
| Thermal conductivity, k [$Wm^{-1}K^{-1}$] | 18 | 1 | 1 | 1 | 0.74 |
| Specific heat, $C_p$ [$JKg^{-1}K^{-1}$] | 339 | 741 | 529 | 385 | 368 |
| Density, ρ [$Kgm^{-3}$] | 5270 | 2221.5 | 2793.6 | 3945.1 | 4640 |

### 2.2 Framework of simulation

A three-dimensional numerical model of the proposed $MgXS_3$ photovoltaic devices was implemented in COMSOL Multiphysics, employing the Semiconductor Module to evaluate its coupled optical and electrical performance. The model accounts for the fundamental physical mechanisms in a solar cell, including light absorption, photogeneration of electron–hole pairs, charge-carrier transport, recombination pathways, and carrier extraction at the electrodes. Within the Semiconductor Module, these processes are described by the self-consistent solution of Poisson's equation together with the steady-state drift–diffusion carrier transport equations for electrons and holes, including radiative and non-radiative recombination channels. By solving this coupled system under illumination, the current–voltage characteristics, power conversion efficiency, and spectral quantum efficiency of the device can be predicted [41-42]. The incident light was modeled as AM 1.5G plane-wave irradiation over the wavelength range of 305–4045 nm.

In the drift–diffusion framework, the photovoltaic response is obtained from the coupled solution of the electrostatic potential ϕ, electron concentration n, and hole concentration p as functions of position. These quantities satisfy Poisson's equation and the carrier continuity equations [41]:

$$\nabla.(\varepsilon_0.\varepsilon_r.\nabla\Phi) = -\rho \tag{1}$$

$$\frac{\delta n}{\delta t} = \frac{1}{q}\nabla j_n + G_n - U_n \tag{2}$$

$$\frac{\delta p}{\delta t} = \frac{1}{q}\nabla j_p + G_p - U_p \tag{3}$$

where, q is the elementary charge, $\varepsilon_0$ and $\varepsilon_r$ are the vacuum and relative permittivity, $G_n$ and $G_p$ are the generation rates of electrons and holes and $U_n$ and $U_p$ are the recombination rates. The space charge density is written as:

$$\rho = q(n - p + N_A - N_D) \tag{4}$$

with $N_A$ and $N_D$ denoting the densities of ionized acceptors and donors, which represent intentional dopants or fixed impurity charges. The electron and hole current densities $J_n$ and $J_p$ include both drift and diffusion contributions and are expressed as [41]:

$$J_n = -q\mu_n n\nabla\Phi + qD_n\nabla n \tag{5}$$

$$J_p = -q\mu_p p\nabla\Phi + qD_p\nabla p \tag{6}$$

where, $\mu_n$ and $\mu_p$ are the mobilities and $D_n$ and $D_p$ are the diffusion coefficients. The latter satisfy the Einstein relations:

$$D_n = \mu_n\frac{K_BT}{q} \tag{7}$$

$$D_p = \mu_p\frac{K_BT}{q} \tag{8}$$

Here, $k_B$ denotes the Boltzmann constant and T is the absolute temperature. The total current density is given by $J = J_n + J_p$. In the simulations, the Semiconductor Module uses user-defined material parameters and includes trap-assisted Shockley–Read–Hall (SRH) recombination via the electron and hole recombination terms $U_n$ and $U_p$. Under steady-state operation, the time-derivative terms in Eqs. (4) and (5) are neglected by setting them to zero.

In this work, the optical generation profile was treated in a standard, physics-based manner and then coupled to the electrical model in COMSOL. Specifically, the depth-dependent total photogeneration rate, $G_{tot}(z)$, was first obtained from the wavelength-resolved generation inside each semiconductor layer. Near-band-edge absorption was modeled with a direct-transition Tauc-type absorption coefficient, and photon attenuation in the absorber was described using the Beer–Lambert law. The resulting wavelength-dependent generation was integrated over the AM 1.5G spectrum (305–4045 nm) to obtain the depth-dependent total generation profile, $G_{tot}(z)$. This profile was finally imported into the Semiconductor Module as the carrier generation term, applied equally to electrons and holes, so that the electrical solution self-consistently reflects the optically generated carrier distribution. This optical-to-electrical coupling procedure follows the methodology reported in the referenced studies [43-45].

To evaluate the thermal response and the temperature gradient of the proposed devices, the relevant thermal outputs were obtained from the Heat Transfer in Solids Module coupling to the Semiconductor Module. The general heat equation can be expressed by partial differential equation and must be solved in steady-state is as follows for a solid body [44]:

$$-k\nabla^2 \mathrm{T} + Q = \rho_p C_p \ \frac{dT}{dt} \tag{9}$$

where k, $C_P$ and $\rho_p$ are the thermal conductivity of the material as a function of temperature [$Wm^{-1}K^{-1}$], the specific heat [$JKg^{-1}K^{-1}$], and the density at constant pressure, respectively. The dominant volumetric loss mechanisms considered here are nonradiative recombination heating and Joule heating within the semiconductor stack. In this framework, Shockley–Read–Hall and Auger recombination rates are determined from the drift–diffusion solution, and the associated energy released to the lattice is represented through a recombination heat-source term of the form [42,44,46]:

$$H_{SRH} = (E_g + 3k_BT) \times (U_{SRH} + U_{Aug}) \tag{10}$$

where, $E_g$ is the bandgap energy, $k_B$ is the Boltzmann constant, T is the local temperature, and $U_{SRH}$ and $U_{Aug}$ are the SRH and Auger recombination rates per unit volume, respectively. Joule heating is included via:

$$H_{Joule} = J \times E \quad (11)$$

where, J is the total current density and $E = -\nabla\phi$ is the electric field and the thermal simulation was performed by Heat Transfer in Solids module by loading this heat sources.

## 2.3 Strategy for meshing

The three-dimensional multilayer device geometry was discretized in COMSOL Multiphysics (version 6.3) using a user controlled meshing sequence to ensure consistent numerical accuracy for all $MgXS_3$ absorbers while keeping the same window and BSF layers. A structured approach was selected because the device is composed of planar stacked layers where strong spatial gradients in electrostatic potential and carrier concentrations are expected near heterointerfaces and contacts. In practice, the meshing procedure applied a Size control, followed by a Mapped mesh on the selected faces and a Swept mesh through the thickness of each layer. This mapped–swept approach aligns elements across interfaces and is well suited to layered semiconductor structures because it enhances numerical stability and minimizes discretization errors in drift–diffusion simulations [47].

The same meshing settings were applied to all three devices, that is, for the $MgTiS_3$, $MgZrS_3$, and $MgHfS_3$ absorbers while keeping identical for window and BSF layers. The mesh density was controlled by setting the maximum element size to 19.8 nm and the minimum element size to 1.65 nm in each model. A maximum element growth factor of 1.08 was imposed to avoid abrupt changes in element size between neighboring regions, and a curvature factor of 0.3 was used to maintain adequate refinement in regions where geometric curvature requires smaller elements. To refine the resolution along the thickness direction, separate swept distributions were applied to the window, absorber, and BSF domains. The Mapped type mesh was defined using 5 elements in the window, 50 elements in the absorber, and 10 elements in the BSF, with element ratios of 2 for the window and 5 for both the absorber and BSF to provide enhanced discretization near the junction regions. Symmetric distribution was applied where appropriate to preserve uniformity across opposite faces of the layered stack.

Mesh quality was evaluated using standard COMSOL quality measures with emphasis on skewness and element size growth [48]. The final mesh consisted of more than 4,000 hexahedral elements and approximately 2,500 quadrilateral elements, resulting in near-perfect element quality and ensuring accurate and stable numerical solutions. Skewness measures how closely each mesh element resembles an ideal reference shape, with lower skewness generally associated with more stable and accurate solutions. The element growth metric describes how gradually element sizes vary between neighboring regions, and values near unity indicate smoother transitions that typically aid convergence and reduce discretization error. These criteria were used to confirm that the adopted mesh is sufficiently fine and well-conditioned for reliable extraction of photovoltaic parameters from the simulated current density voltage characteristics.

## 3. Result and discussion

### 3.1 Generation rate

The depth-dependent total photogeneration profile, $G_{tot}(z)$, has been computed for three device variants that differ only in the $MgXS_3$ absorber composition (X = Ti, Zr, Hf), while the remaining stack is kept identical. Each three-dimensional structure comprises a ZnSe window layer, an $MgXS_3$ absorber, and an $Sb_2S_3$ back-surface-field (BSF) layer. Fig. 2 presents the corresponding 3D distributions of $G_{tot}$ for $MgTiS_3$, $MgZrS_3$, and $MgHfS_3$ perovskites. For all absorbers, photogeneration is highest near the ZnSe/$MgXS_3$ junction because the incident photon flux first reaches the active region at this interface, and it progressively decreases toward the rear side as the optical intensity is attenuated along the thickness direction. Quantitatively, the $MgTiS_3$ device exhibits the largest peak generation rate at the ZnSe/$MgTiS_3$ interface, reaching approximately $1.93 \times 10^{22}$ cm$^{-3}$ s$^{-1}$. For $MgZrS_3$, the maximum $G_{tot}$ decreases to about $1.38 \times 10^{22}$ cm$^{-3}$ s$^{-1}$, while $MgHfS_3$ shows the lowest peak value of roughly $1.11 \times 10^{22}$ cm$^{-3}$ s$^{-1}$. The absorber bandgaps are 1.10 eV for $MgTiS_3$, 1.30 eV for $MgZrS_3$, and 1.43 eV for $MgHfS_3$, and the systematic reduction in peak photogeneration from $MgTiS_3$ to $MgHfS_3$ is consistent with the increasing bandgap, which narrows the absorbed spectral range and reduces the number of above-band-gap photons contributing to carrier generation. Since the $MgXS_3$ layer absorbs the majority of photons with energies above its bandgap, the $Sb_2S_3$ BSF makes only a minor contribution to the overall generation. Although the three devices exhibit similar qualitative trends, the absolute magnitude and the decay rate of $G_{tot}(z)$ vary with absorber bandgap, reflecting differences in spectral absorption and photon penetration depth.

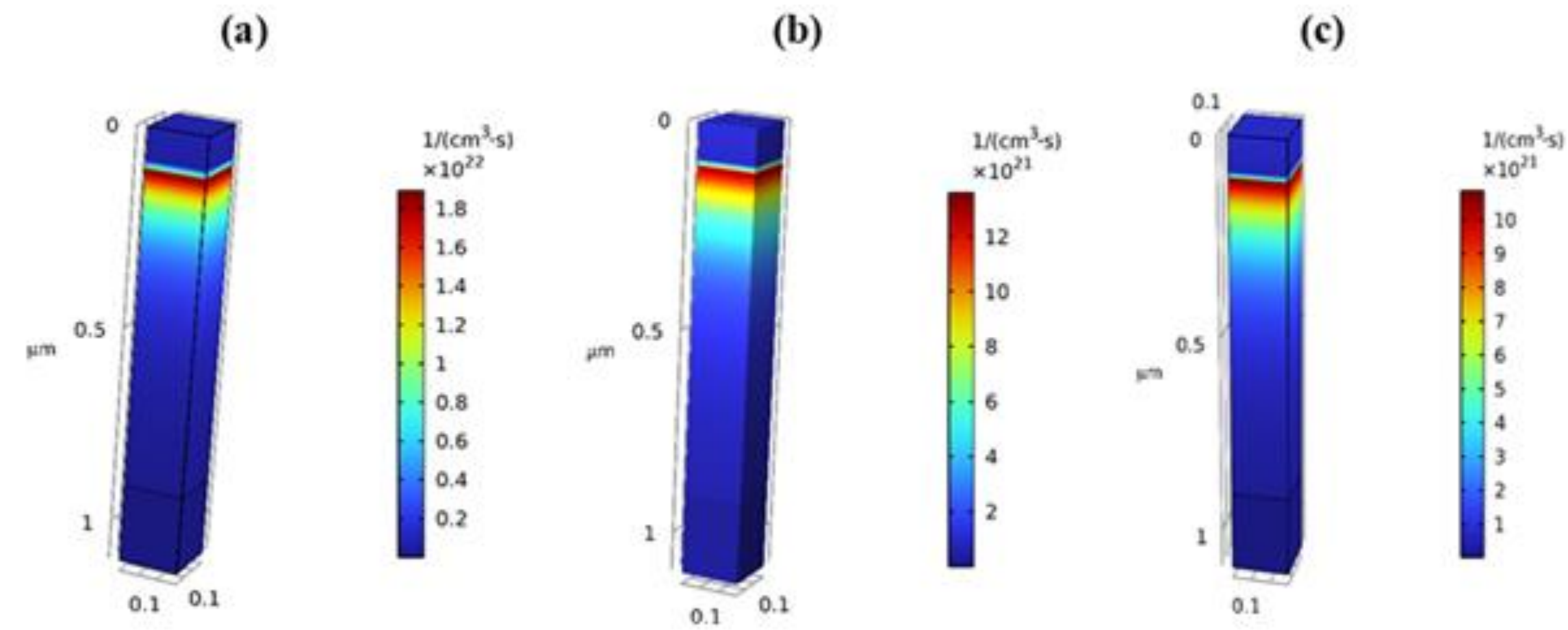

**Fig. 2:** Generation profile of the proposed devices: (a) $MgTiS_3$, (b) $MgZrS_3$ and (c) $MgHfS_3$ in 3D structure.

### 3.2 Effect of absorbers on PV parameters of the proposed devices

Fig. 3(a) demonstrates the impacts of thickness of $MgXS_3$ (X=Ti, Zr, Hf) absorbers on the n-ZnSe/p-$MgXS_3$/p$^+$-$Sb_2S_3$ perovskite solar cells. The optimized values of absorber thickness, doping concentration and defect density are 800 nm, $10^{17}$ cm$^{-3}$ and $10^{14}$cm$^{-3}$, respectively. The $J_{SC}$ increases steadily from 39.44 mA/cm$^2$ to 43.53 mA/cm$^2$ for $MgTiS_3$, 31.61 to 35.97 mA/cm$^2$ for $MgZrS_3$, 27.29 to 31.31 mA/cm$^2$ for $MgHfS_3$ with the increment of thickness of $MgXS_3$ absorber from 400 nm to 1200 nm. FF also increases from 84.43% to 84.63% for $MgTiS_3$, 86.12% to 87.21% for $MgZrS_3$ and 87.11% to 88.43% in case of $MgHfS_3$ solar cell. Consistent with these trends, the PCE increases with thickness, reaching optimized condition for $MgTiS_3$ is 26.72%, 28.18% for $MgZrS_3$ and 28.16% for $MgHfS_3$ devices. These improvements are mainly optical in origin, as increasing absorber thickness enhances photon absorption, raises photogeneration and $J_{SC}$, and typically yields a modest increase in FF [49]. On the other hand, $V_{OC}$ decreases steadily starting from 0.76 V and ending at 0.73 V for $MgTiS_3$, 0.96 V to 0.93 V for $MgZrS_3$, 1.09V to 1.06V for $MgHfS_3$ devices for aforementioned absorber thickness because the larger active volume elevates bulk Shockley–Read–Hall recombination, raising the dark current ($J_0$ ) and thereby suppressing $V_{OC}$ despite improved light harvesting [50].

In Fig. 3(b), the effect of doping in $MgXS_3$ (X=Ti, Zr, Hf) absorbers on the photonic device performance has been shown with the variation from $10^{15}$ $cm^{-3}$ to $10^{19}$ $cm^{-3}$ for fixed width of absorber at 800 nm. Here, $V_{OC}$ rises from 0.7 V to 0.85 V for $MgTiS_3$, 0.89 V to 1.08 V in case of $MgZrS_3$ and 1.02 V to 1.19 V for $MgHfS_3$ cells. Very small changes have been observed in case of $J_{SC}$ for $MgTiS_3$, $MgZrS_3$ and $MgHfS_3$ devices. FF enhances steadily from 77.544% to 87.12% for $MgTiS_3$, 82.25% to 89.54% for $MgZrS_3$ and 82.07% to 89.16% for $MgHfS_3$. Similarly, PCE also increases from 23.2% to 31.67% for $MgTiS_3$, 25.24% to 33.24% for $MgZrS_3$ and in case of $MgHfS_3$ PCE rises from 25.09% to 31.78%. Increasing acceptor doping in the p-type absorber strengthens the built-in potential and junction field, lowering the minority-carrier concentration and suppressing the dark current, thereby enhancing $V_{OC}$. At moderate doping, efficient depletion-field collection combined with reduced Shockley–Read–Hall recombination drives the effective diode ideality factor toward unity and improves the FF. However, at excessive doping, trap-assisted recombination elevates $J_0$, which in turn reduces both $V_{OC}$ and the FF [50-51]. The insensitivity of $J_{SC}$ to doping over $10^{15}$ $cm^{-3}$ to $10^{19}$ $cm^{-3}$ reflects that, in thin- film absorbers whose diffusion length exceeds their thickness, the short-circuit current is mainly controlled by photogeneration and transport, and it changes little with background doping unless doping is high enough to increase recombination [52].

Fig. 3(c) displays the influence of defect density of $MgXS_3$ (X=Ti, Zr, Hf) absorbers on the photovoltaic device performance has been investigated with the variation from $10^{12}$ $cm^{-3}$ to $10^{16}$ $cm^{-3}$ for fixed absorber width of 800 nm. Here, both $V_{OC}$ and PCE decrease with the increment of defect density. PCE reduces from 32.41 % to 17.71% for $MgTiS_3$, 32.81% to 19.86% in case of $MgZrS_3$ and 31.16% to 19.58% for $MgHfS_3$ and $V_{OC}$ reduces from 0.94 V to 0.61V for $MgTiS_3$, 1.12 V to 0.8 V for $MgZrS_3$ and 1.17 V to 0.94 V for $MgHfS_3$ devices. These trends are consistent with enhanced trap-assisted Shockley–Read–Hall recombination that shortens carrier lifetimes and elevates the dark saturation current, thereby limiting $V_{OC}$ and overall efficiency [49]. $J_{SC}$ is initially remains relatively stable because optical absorption is unchanged and the diffusion length exceeds the film thickness at low defect densities [52], but it decreases with further defect introduction, from 42.74 $mA/cm^2$ to 39.01 $mA/cm^2$ in case of $MgTiS_3$, 34.5 $mA/cm^2$ to 32.04 $mA/cm^2$ for $MgZrS_3$ and 29.98 mA/$cm^2$ to 26.09 mA/$cm^2$ for $MgHfS_3$. However, FF fluctuates both for $MgTiS_3$ and $MgZrS_3$ reflecting changes in the apparent diode ideality factor and the dominant recombination path evolve with defect density, while series and shunt elements shift the operating point [53], and decreases monotonically for $MgHfS_3$ cell.

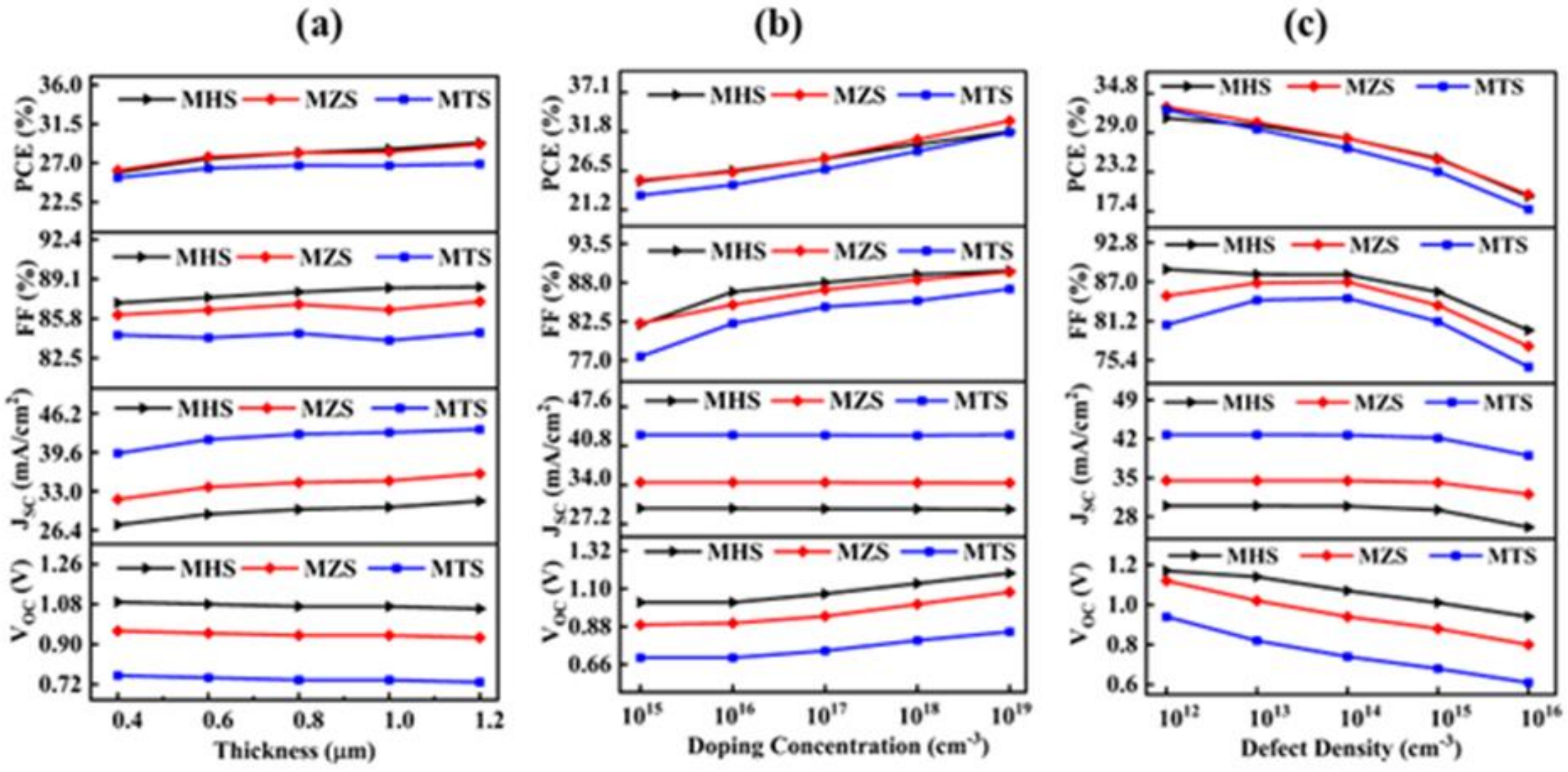

**Fig. 3:** The effects of changing absorber's: (a) thickness, (b) doping and (c) defects on the $MgXS_3$ cell.

Overall, Fig. 3 indicates that the $MgXS_3$ absorbers fall within a bandgap range that is favorable for high-efficiency single-junction photovoltaics. $MgZrS_3$ and $MgHfS_3$ generally yield marginally higher optimized efficiencies than $MgTiS_3$ when the device configuration is otherwise unchanged. This behavior follows the well-known single-junction band-gap trade-off, in which the narrower-gap $MgTiS_3$ can harvest more long-wavelength photons and produce a higher $J_{SC}$, but it typically exhibits a reduced $V_{OC}$ because a smaller band gap increases thermalization losses and strengthens recombination-driven voltage penalties. In contrast, the wider-bandgap $MgZrS_3$ and $MgHfS_3$ absorbers provide higher photovoltage while maintaining sufficient current generation, which leads to a modest net advantage in PCE within the optimized n-ZnSe/p-$MgXS_3$/$p^+$-$Sb_2S_3$ device architecture.

### 3.3 Effect of window on PV parameters of the proposed device

Fig. 4 examines how modifications to the ZnSe layer's physical characteristics influence the performance of $MgXS_3$ (X = Ti, Zr, Hf)–based PV cells. Fig. 4(a) shows the impact of increasing the ZnSe thickness from 100 nm to 500 nm. The $V_{OC}$ remains essentially unchanged, varying only from 0.73 to 0.74 V for $MgTiS_3$, 0.93 to 0.94 V for $MgZrS_3$, and 1.06 to 1.07 V for $MgHfS_3$ devices. This weak dependence arises because the minority carrier diffusion length is much larger than the ZnSe thickness, so recombination in the window layer is negligible [49]. In contrast, $J_{SC}$ and FF exhibit only small fluctuations over this thickness range for all three absorbers. The efficiency decreases slightly from 26.72% to 25.915% for $MgTiS_3$, from 28.177% to 27.25% for $MgZrS_3$, and from 28.16% to 27.359% for $MgHfS_3$ cells. This behavior can be linked to increased effective series resistance and reduced shunt resistance as the window layer thickens, which together introduce additional resistive and leakage losses and slightly degrade device performance [54].

Fig. 4(b) shows the influence of ZnSe window doping, varied from $10^{15}$ $cm^{-3}$ to $10^{19}$ $cm^{-3}$, on the performance of the devices at a fixed window thickness of 100 nm. The $V_{OC}$ decreases for all three absorbers, falling from 0.78 V to 0.60 V for $MgTiS_3$, from 0.98 V to 0.76 V for $MgZrS_3$, and from 1.10 V to 0.94 V for $MgHfS_3$. The $J_{SC}$ remains relatively constant as the doping level growths from $10^{15}$ $cm^{-3}$ to $10^{17}$ $cm^{-3}$ for each material, then shows a slight decline at higher doping. The FF decreases gradually in all cases, from 86.18% to 81.55% for $MgTiS_3$, from 88.07% to 84.48% for $MgZrS_3$ and from 89.02% to 86.30% for $MgHfS_3$ devices. As a result, the PCE also decreases monotonically with increasing window doping. These trends can be ascribed to the fact that higher carrier concentrations enhance recombination, particularly Shockley–Read–Hall processes in and near the window–absorber interface, which increases the dark saturation current and reduces both $V_{OC}$ and FF. At high doping, additional recombination and field screening effectively raise the impact of parasitic resistances and lead to observable losses in $J_{SC}$, FF and PCE [49, 54].

In Fig. 4(c), the influence of defect density on the performance of device has been displayed. Here the value of defect density is set to a range beginning from $10^{12}$ $cm^{-3}$ to $10^{16}$ $cm^{-3}$ and window's thickness is fixed as 100 nm. Over this range, all key parameters ($V_{OC}$, $J_{SC}$, FF and PCE) remain essentially unchanged, indicating that carrier lifetime and diffusion length in the window layer are sufficiently large compared with its thickness, so recombination in ZnSe has a negligible impact on device operation. Beyond this range, however, a further increase in defect density would be expected to enhance recombination, shorten diffusion lengths and hinder carrier transport, ultimately degrading overall device performance [49,55].

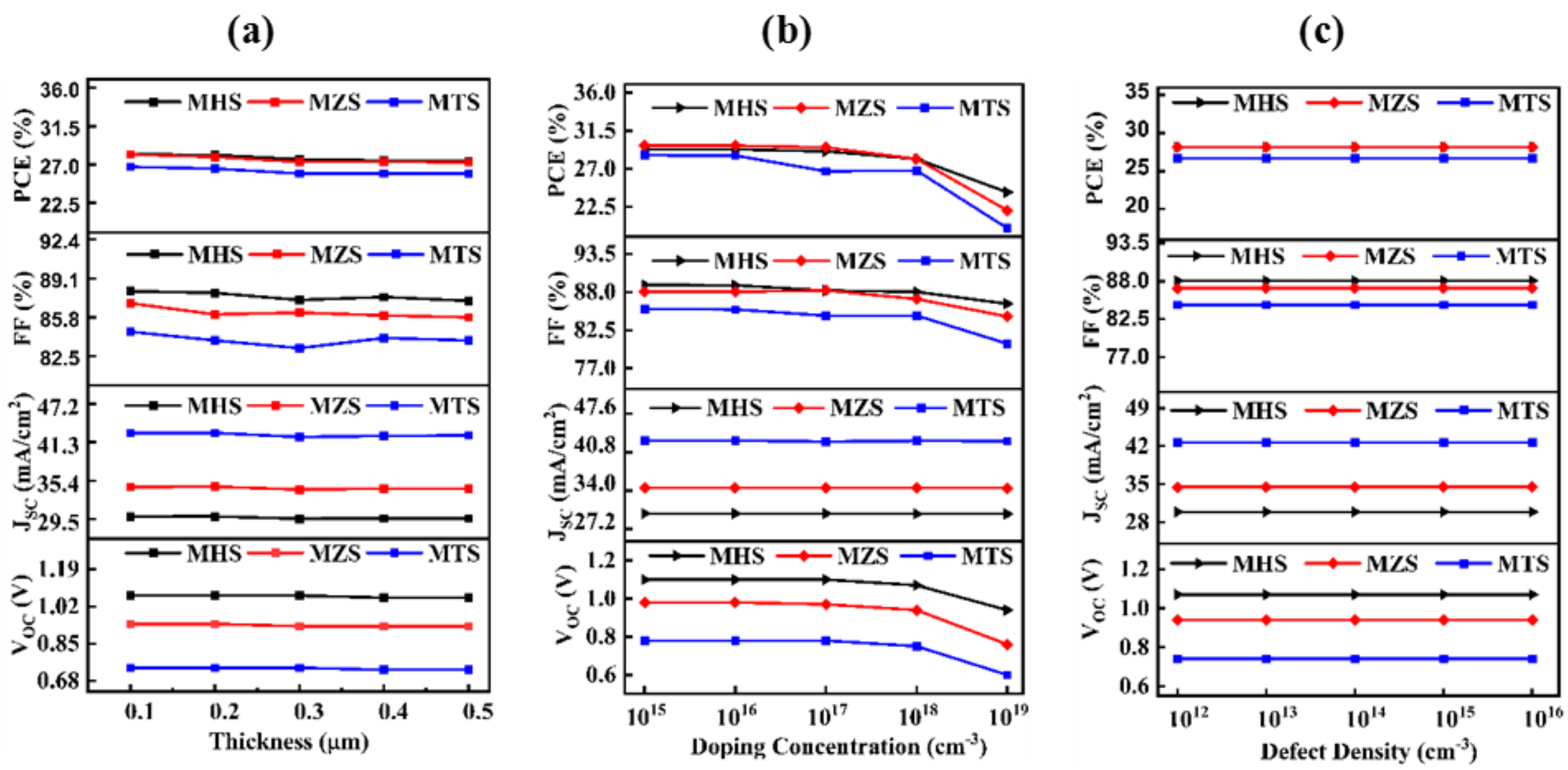

**Fig. 4:** The effects of changing ZnSe window's: (a) thickness, (b) doping and (c) defects on the $MgXS_3$ cells.

### 3.4 Effect of $Sb_2S_3$ BSF layer on PV parameters of the proposed devices

In this part, the impacts of $Sb_2S_3$ BSF's layer on the performance of offered cell have been investigated. Fig. 5(a) shows the effects of changing BSF's thickness inaugurating from 100 nm to 500 nm on various parameters of PV device. It has been seen that $V_{OC}$ is completely constant for $MgTiS_3$ (0.74 V), $MgZrS_3$ (0.94 V) and $MgHfS_3$ (1.07 V) devices, indicating the charge-carrier lifetime and diffusion length in the absorber remain sufficiently large to offset recombination within the BSF layer [51]. The $J_{SC}$ decreases slightly from 100 to 400 nm and then recovers at 500 nm for all three absorbers, reflecting a trade-off between added rear-side transport and recombination losses in a thicker BSF and improved back-surface passivation and carrier collection when the BSF becomes sufficiently thick [56]. The fill factor changes only weakly, with a minor decrease for $MgTiS_3$ and slight increases for $MgZrS_3$ and $MgHfS_3$, confirming that the overall device performance is only weakly sensitive to BSF thickness within the investigated window.

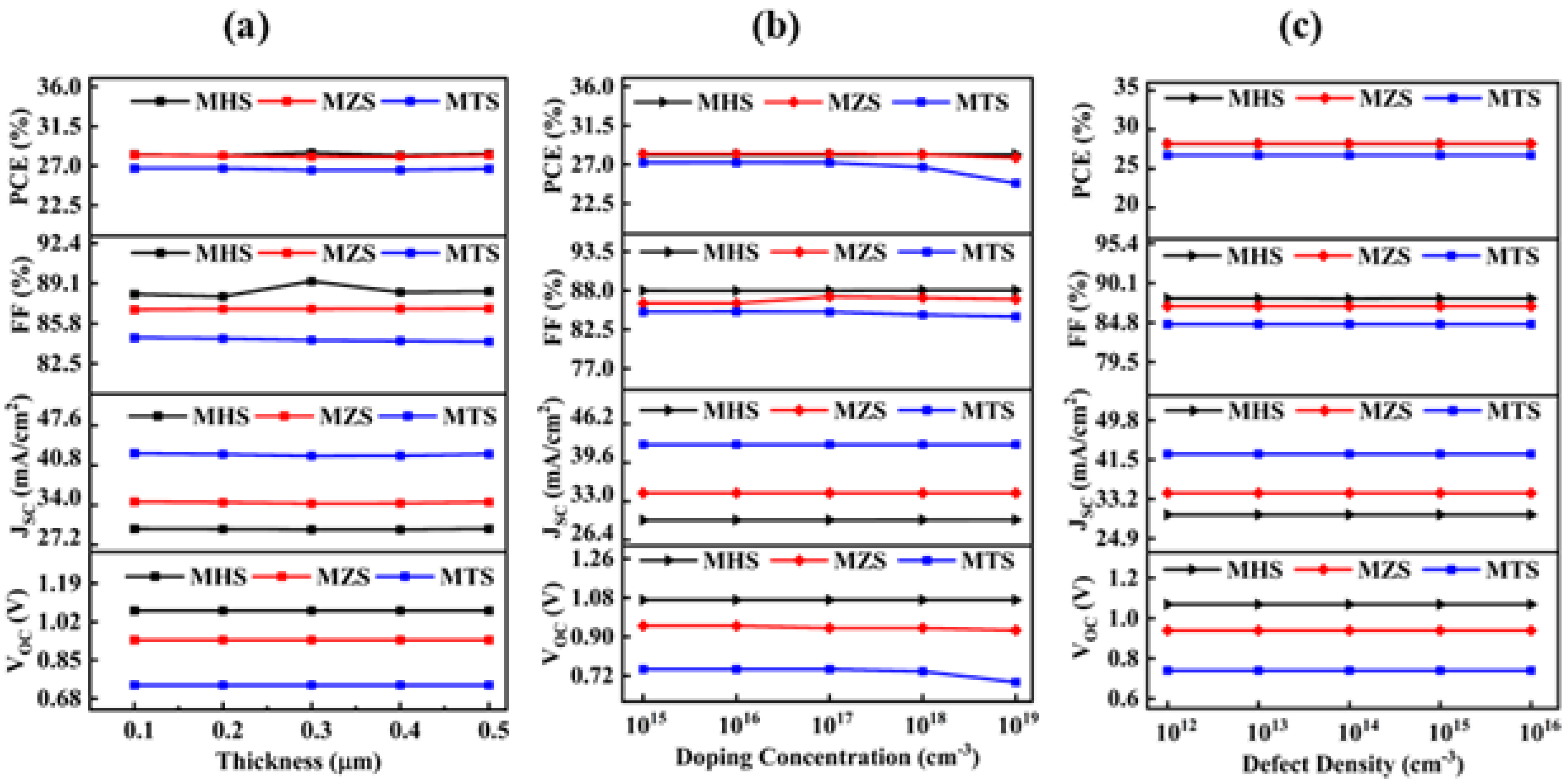

**Fig. 5:** The effects of changing the $Sb_2S_3$ BSF's: (a) thickness, (b) doping and (c) defects on the $MgXS_3$ PV cell.

Fig. 5(b) presents the effect of BSF doping concentration, varied from $10^{15}$ to $10^{19}$ $cm^{-3}$ while keeping the BSF thickness fixed at 200 nm. A trend similar to the window-layer doping sweep is observed. For all three absorbers, $V_{OC}$ decreases slightly with increasing doping, whereas $J_{SC}$ remains nearly unchanged. In addition, both FF and PCE show a clear downward trend across the full doping range.

Fig. 5(c) shows the effect of defect density, varied from $10^{12}$ $cm^{-3}$ to $10^{16}$ $cm^{-3}$, on device performance while keeping the BSF thickness fixed at 200 nm. Similar to the window-layer defect-density sweep, all key photovoltaic parameters remain nearly constant across the investigated range for all three absorbers.

### 3.5 Effect of working temperature and resistances on PV parameters

#### *3.5.1. Influence of working temperature*

Operating temperature is a decisive parameter for photovoltaic (PV) operation because it directly affects both output performance and device reliability. Evaluating solar cells under elevated thermal conditions is therefore important for assessing operational stability, since a temperature rise can induce interfacial stress and structural mismatch between stacked layers, which may translate into electrical instability in many PV configuration [57]. In this study, the temperature dependence of the proposed $MgXS_3$ (X = Ti, Zr, Hf) devices has been examined over 200–500 K, as illustrated in Fig. 6(a). A clear and monotonic reduction in the open-circuit voltage ($V_{OC}$) is observed with increasing temperature. Specifically, $V_{OC}$ decreases from 0.86 to 0.50 V for $MgTiS_3$, from 1.07 to 0.70 V for $MgZrS_3$ and 1.20 to 0.8 V for $MgHfS_3$ solar cells, as the temperature increases from 200 to 500 K. This behavior is mainly attributed to thermally activated recombination and the accompanying increase in reverse saturation current density ($J_0$), which reduces the quasi-Fermi level splitting. Consistent with diode theory, the inverse dependence of $V_{OC}$ on $J_0$ explains why $V_{OC}$ degrades as temperature rises [49,58].

In contrast, the $J_{SC}$ exhibits only a slight upward trend for all three absorbers across the investigated range. This marginal increase can be linked to a reduction in the effective bandgap at higher temperatures [59], which enhances absorption near the long-wavelength edge and enables additional photon harvesting, thereby generating more charge carriers.

The fill factor (FF), however, declines nearly linearly with temperature, indicating progressive losses in charge transport and junction quality. At 200 K, FF values of 91.49%, 91.48%, and 89.82% are obtained for $MgHfS_3$, $MgZrS_3$, and $MgTiS_3$ cells, respectively. These values drop to 78.93%, 75.99%, and 71.27% at 500 K. The deterioration in FF can be explained by increased recombination at elevated temperature together with stronger resistive losses, effectively higher series resistance and lower shunt resistance, which reduce the squareness of the J–V characteristics and drive the diode response away from ideal behavior [54].

As expected, the PCE follows the combined temperature response of $V_{OC}$, $J_{SC}$, and FF. The highest efficiencies are achieved at 200 K, reaching 32.8% for $MgHfS_3$, 33.45% for $MgZrS_3$, and 33.01% for $MgTiS_3$, but they decrease substantially to 18.78%, 18.27%, and 15.16%, respectively, at 500 K. Although $J_{SC}$ increases slightly with temperature, the pronounced reduction in $V_{OC}$ and FF dominates the overall outcome, leading to a continuous decline in efficiency. Overall, these results indicate that thermal effects, particularly the rise of $J_0$ and recombination-driven voltage losses, can significantly limit device performance and may accelerate degradation under practical operating conditions. Therefore, maintaining lower operating temperatures and implementing effective thermal management are crucial for preserving high efficiency [60].

### *3.5.2. Influence of series and shunt resistances*

Series resistance ($R_S$) and shunt resistance ($R_{Sh}$) are two key parasitic elements that largely determine the practical J–V behaviors of a solar cell and therefore its extractable power. In an ideal device, the current density follows the Shockley diode relation under one-sun illumination, but real cells deviate from ideality because part of the generated voltage is lost through internal resistive drops and part of the photogenerated current is lost through leakage pathways [61]. This behavior is commonly described by the single-diode model including $R_S$ and $R_{Sh}$, which can be written as [62]:

$$I = I_L\text{-}I_0 \exp\left[\frac{q(V+IR_S)}{nK_BT}\right]\text{-}\frac{V+IR_S}{R_{SH}} \tag{12}$$

Here, $I_L$ is the light-generated current, $I_0$ is the reverse saturation current, q is the elementary charge, n is the diode ideality factor, $K_B$ is the Boltzmann constant, and T is the absolute temperature. In this framework, $R_S$ represents the cumulative resistive losses associated with the absorber bulk, interfaces, transport layers, and metal contacts, while $R_{Sh}$ captures leakage channels arising from defects, pinholes, and recombination-assisted current paths [49,54].

As presented in Fig. 6(b), increasing the series resistance ($R_S$) produces a negligible change in the open-circuit voltage ($V_{OC}$), which remains essentially constant for all three absorbers. This behaviour is expected because, under open-circuit conditions, the net current is zero and the voltage drop across $R_S$ becomes insignificant, so $R_S$ has minimal influence on $V_{OC}$. In contrast, the short-circuit current density ($J_{SC}$) shows a slight decrease as $R_S$ increases, because the additional resistive loss impedes charge transport and reduces the effective current that can be extracted under load [63]. Consequently, both the fill factor (FF) and the power conversion efficiency (PCE) decline markedly with increasing $R_S$. When $R_S$ is increased from 0 to 10 $\Omega.cm^2$, FF decreases from 87.8% to 62.32% for $MgHfS_3$, from 86.18% to 54.16% for $MgZrS_3$, and from 83.84% to 38.84% for $MgTiS_3$ cells. Over the same $R_S$ range, PCE drops from 28.16% to 19.99% for $MgHfS_3$, from 28.18% to 17.70% for $MgZrS_3$, and from 26.72% to 12.37% for $MgTiS_3$. These trends confirm that series resistance primarily degrades device performance by lowering the FF and limiting power extraction rather than by altering $V_{OC}$.

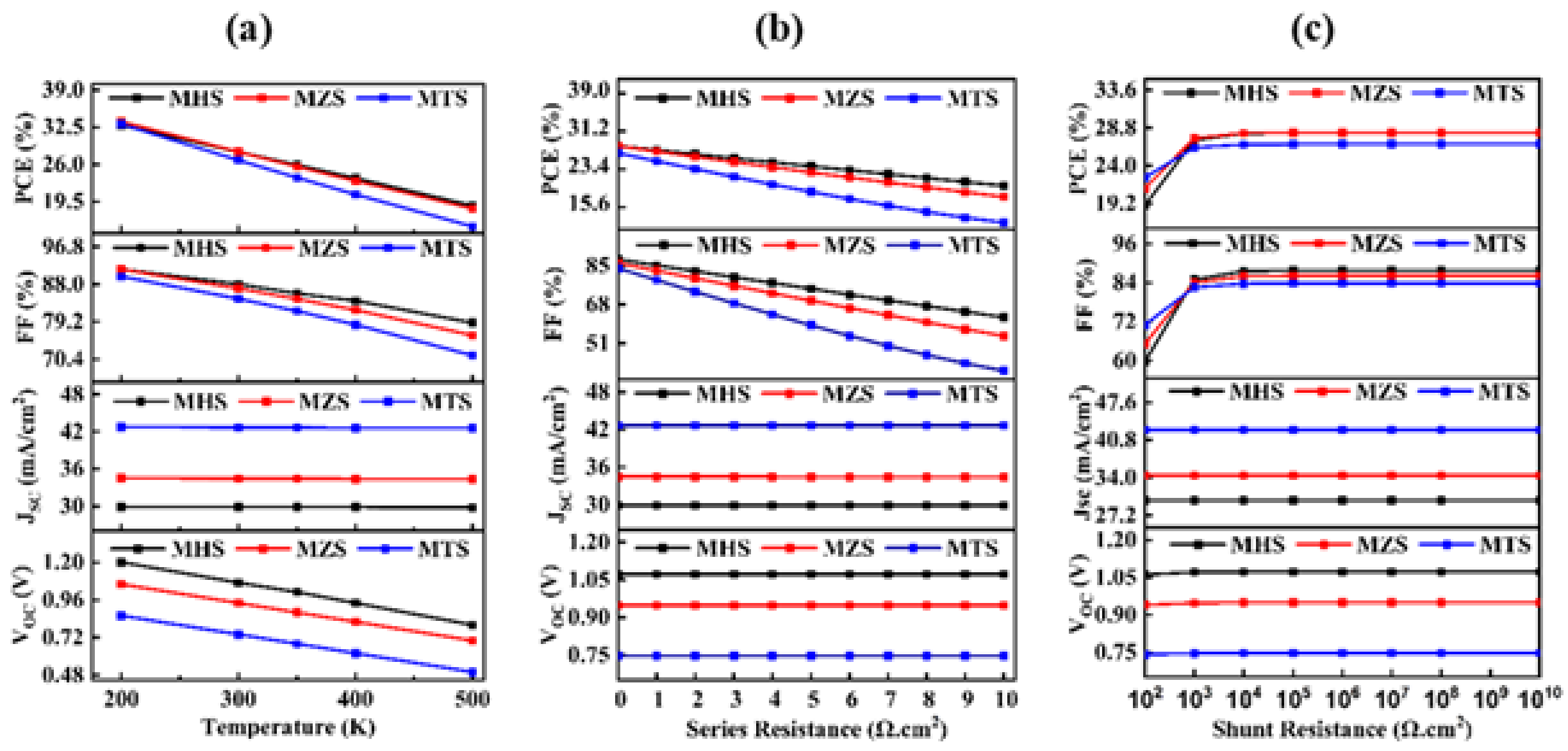

**Fig. 6:** The variation in key performance metrics of the $MgXS_3$ (X = Ti, Zr, Hf)–based photonic devices as a function: (a) temperature, (b) series resistance, and (c) shunt resistance.

In Fig. 6(c), the influence of shunt resistance ($R_{Sh}$) is examined over the range of $1\times10^2$ $\Omega.cm^2$ to $1\times 10^{10}$ $\Omega.cm^2$. The results show that variations in $R_{Sh}$ across this interval have a negligible effect on the short-circuit current density ($J_{SC}$), indicating that photocurrent generation remains nearly unchanged. In contrast, the open-circuit voltage ($V_{OC}$) is sensitive to $R_{Sh}$ at low values, where leakage pathways are pronounced. When $R_{Sh}$ increases from $10^2$ $\Omega.cm^2$ to $10^4$ $\Omega.cm^2$, $V_{OC}$ rises slightly from 1.06 V to 1.07 V for $MgHfS_3$, from 0.94 V to 0.95 V for $MgZrS_3$, and from 0.74 V to 0.75 V for $MgTiS_3$ solar cells. Beyond $10^4$ $\Omega.cm^2$, $V_{OC}$ reaches a saturated value and remains essentially constant for all absorbers. A similar trend is observed for the fill factor (FF) and power conversion efficiency (PCE), which are considerably lower at small $R_{Sh}$ due to strong shunt losses, but become stable once $R_{Sh}$ exceeds approximately $10^4$ $\Omega.cm^2$. This behaviour arises because low $R_{Sh}$ promotes severe leakage current through dominant shunt pathways, which diverts photogenerated carriers away from the external circuit and distorts the diode response [63-64]. Once $R_{Sh}$ becomes sufficiently large, the shunt leakage current becomes negligible compared with the photocurrent and diode current under normal operating conditions. Therefore, when $R_{Sh}$ surpasses a threshold value, typically above $10^4$ $\Omega.cm^2$, further increases provide minimal additional improvement in $V_{OC}$, FF, and PCE [65].

### 3.6 Influence of $Sb_2S_3$ BSF layer on device performance and quantum efficiency

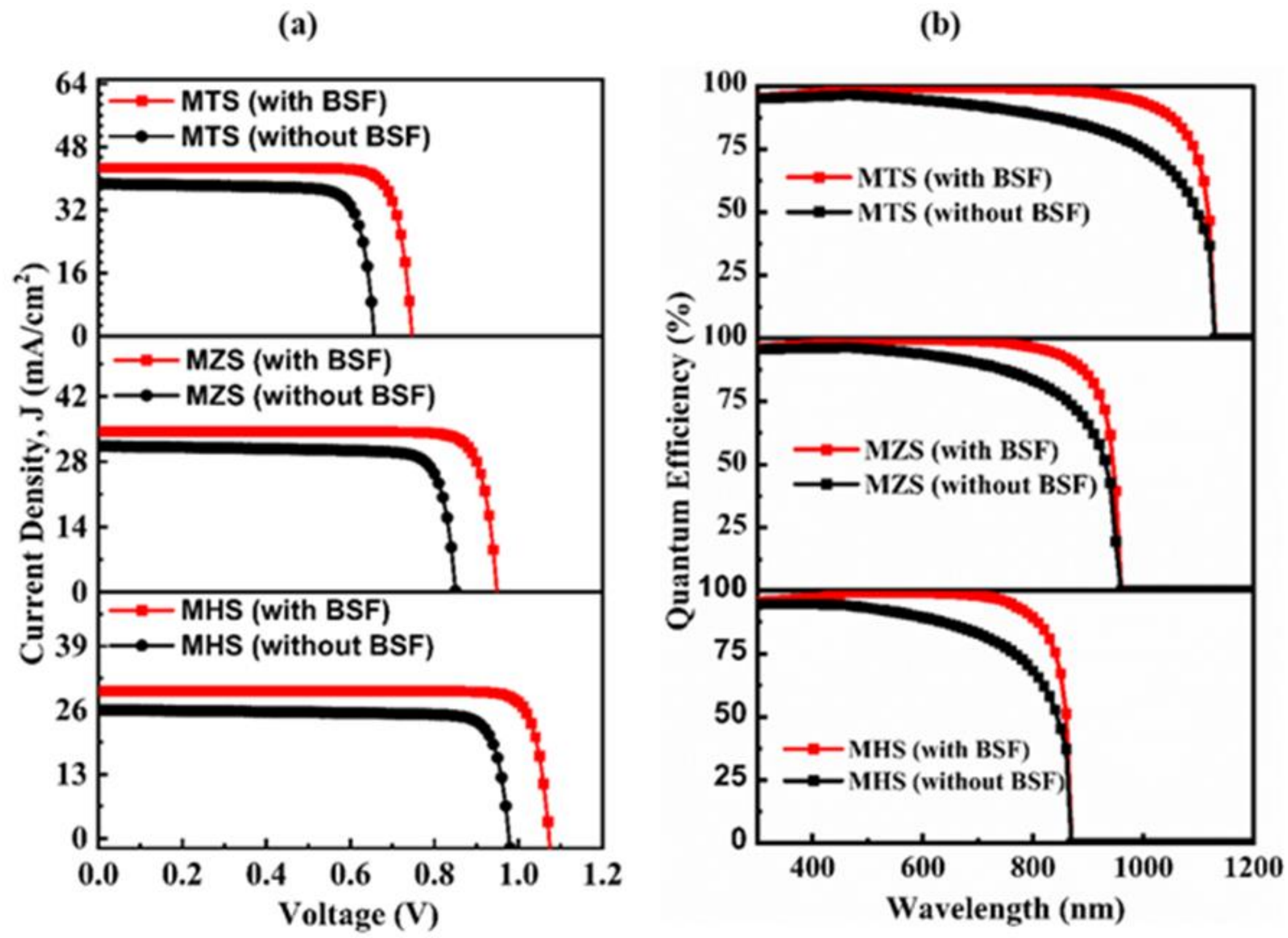

**Fig. 7:** Photovoltaic characteristics of the $MgXS_3$ (X = Ti, Zr, Hf)–based PV cells, including (a) J-V curves and (b) QE, analyzed with and without the incorporation of $Sb_2S_3$ BSF layer.

Fig. 7(a) illustrates the influence of incorporating the $Sb_2S_3$ BSF layer of all $MgXS_3$-based devices through J-V characteristics. In the absence of the BSF layer, the $MgZrS_3$-based device exhibits the highest PCE of 22.025%, with $J_{SC}$, $V_{OC}$, and FF values of 31.322 mA/cm$^2$, 0.85 V, and 82.73%, respectively. The $MgHfS_3$-based device delivers $J_{SC}$, $V_{OC}$, FF, and PCE values of 26.06 mA/cm$^2$, 0.97 V, 85.075%, and 21.505%, respectively, while the $MgTiS_3$-based device shows 38.025 mA/cm$^2$, 0.65 V, 82.29%, and 20.34%, respectively. After introducing the $Sb_2S_3$ BSF layer, all three devices demonstrate enhanced performance. The $MgZrS_3$-based device achieves the highest PCE of 28.18%, with $J_{SC}$, $V_{OC}$, and FF values of 34.459 mA/cm$^2$, 0.94 V, and 86.99%, respectively. Similarly, the $MgHfS_3$-based device reaches 29.894 mA/cm$^2$, 1.07 V, 88.018%, and 28.16%, respectively, whereas the $MgTiS_3$-based device improves to 42.687 mA/cm$^2$, 0.74 V, 84.577%, and 26.72%, respectively. These gains are mainly attributed to the BSF layer, which improves carrier collection near the rear contact and increases the contribution of long-wavelength photons, as reflected in the enhanced quantum-efficiency spectra as shown in Fig. 7(b) [23].

Fig. 7(b) compares the spectral quantum efficiency (QE) of the three $MgXS_3$(X = Ti, Zr, Hf) devices with and without an $Sb_2S_3$ BSF layer. In all cases, the QE spectra display the typical behavior of thin-film p–n junction devices, showing suppressed response at shorter wavelengths, a broad high-QE plateau toward the near-infrared, and a sharp roll-off at the absorption edge. The low QE in the UV–blue region arises because the very short absorption length causes a large fraction of the incident photons to be absorbed in the ZnSe window and in the front surface region, where strong recombination prevents efficient carrier collection [66]. In the near-infrared region, $MgTiS_3$, $MgZrS_3$, and $MgHfS_3$ cells exhibit efficient carrier collection, producing a broad, nearly constant QE plateau. At longer wavelengths, the QE declines rapidly once the photon energy drops below the absorber band gap, since sub-bandgap photons are not absorbed and thus generate no photocurrent.

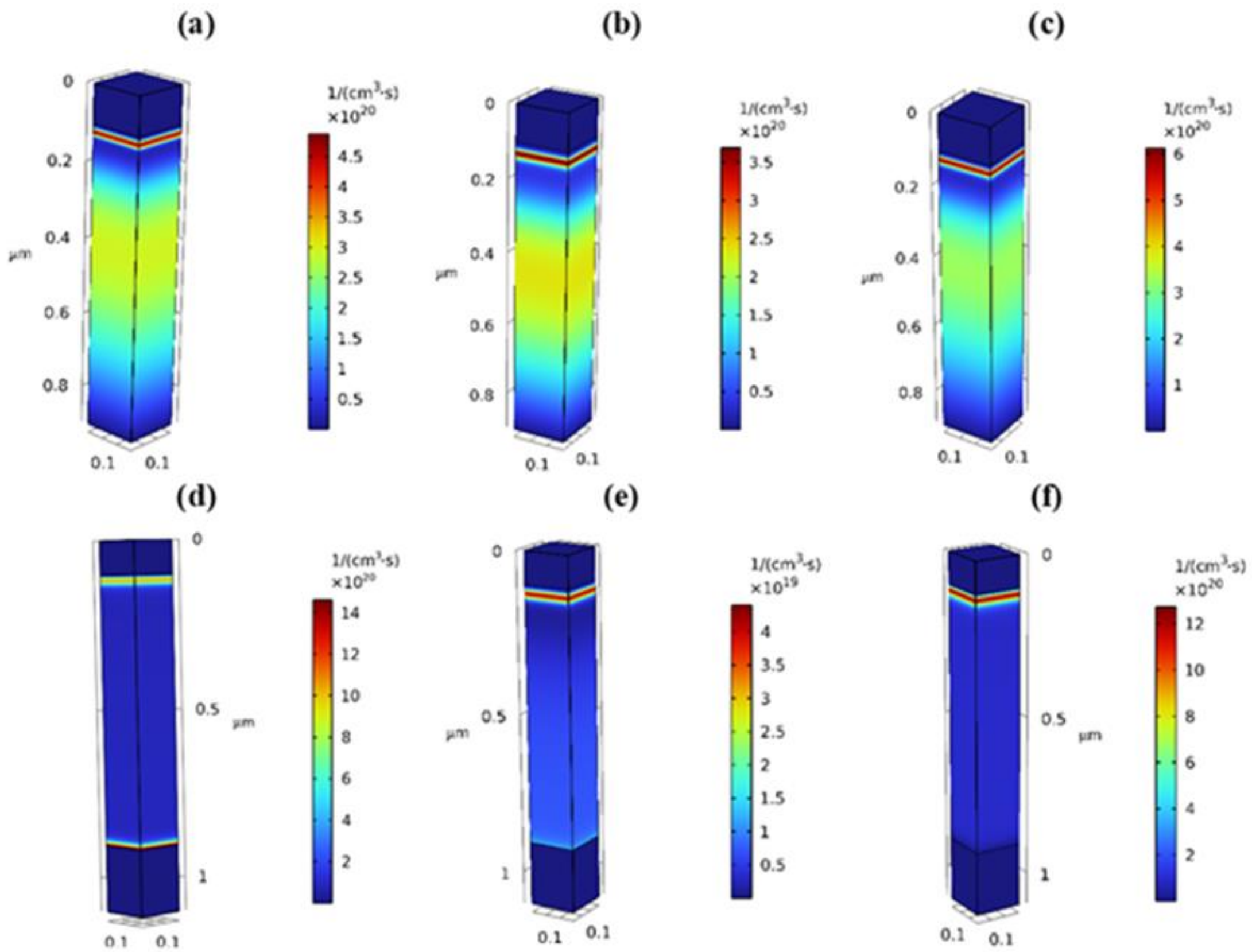

Fig. 8. Recombination-rate distributions of $ZnSe/MgXS_3/Sb_2S_3$ solar cells without BSF for: (a) $MgHfS_3$, (b) $MgZrS_3$, and (c) $MgTiS_3$ PV cells; and with $Sb_2S_3$ BSF fro: (d) $MgHfS_3$, (e) $MgZrS_3$, and (f) $MgTiS_3$ PV cells.

In addition, the recombination profile as shown in Fig. 8 provides further physical insight into the role of the $p^+$-$Sb_2S_3$ BSF layer. Without incorporating the BSF layer, the recombination rate appears an elevated recombination extends over a substantial portion of the absorber region as shown in Fig. 8(a-c). After incorporating the BSF layer as shown in Fig. 8(d-f), the recombination activity becomes confined near the both junction region, while the majority of the absorber bulk volume exhibits considerably lower recombination rates due to a rear electric field at the $MgXS_3/Sb_2S_3$ interface and thus leads to simultaneous increases in $J_{SC}$, $V_{OC}$, and FF [49].

### 3.7 Thermal behavior of the proposed chalcogenide PSCs

In addition to optical absorption and electrical output, thermal behavior is an important aspect of photovoltaic device operation because internally generated heat can alter transport properties, increase recombination activity, and accelerate material and interface degradation during long-term use [67]. Under solar illumination, only a fraction of the absorbed photon energy is converted into useful electrical power, while the remaining portion is dissipated as heat through several loss pathways [42]. In the present devices, the dominant heat sources considered are nonradiative recombination heating, mainly associated with SRH recombination, and Joule heating arising from carrier transport through regions with finite electrical conductivity and interfacial resistances. Spatially resolving these heat-generation terms provides insight into where hot spots form within the multilayer stack and helps identify performance-limiting regions that may require interface passivation, defect control, or improved thermal management.

#### *3.7.1 Spatial distribution of nonradiative recombination heating in the proposed devices*

Fig. 9 presents three-dimensional maps of nonradiative recombination heating for $MgXS_3$(X = Ti, Zr, Hf) absorbers under short-circuit (V = 0 V) and optimum operating bias (Vopt). Panels (a), (b), and (c) correspond to $MgTiS_3$, $MgZrS_3$, and $MgHfS_3$ at V = 0 V, while panels (d), (e), and (f) show $MgTiS_3$, $MgZrS_3$, and $MgHfS_3$ at Vopt, respectively.

In the investigation from Fig. 9(a) to 9(c), the nonradiative (SRH) recombination heating at V = 0 V is primarily concentrated near the front heterojunction at the ZnSe/absorber interface, while the absorber bulk remains comparatively weakly heated. The peak volumetric heat-generation rates reach about $4.78 \times 10^9$ W/m$^3$ for $MgTiS_3$, $3.42 \times 10^9$ W/m$^3$ for $MgZrS_3$, and $2.68 \times 10^9$ W/m$^3$ for $MgHfS_3$ devices. This localization is consistent with short-circuit operation, where the internal field promotes rapid carrier extraction and reduces the steady-state carrier density in the absorber, thereby limiting bulk trap-assisted recombination relative to forward-biased operation [68].

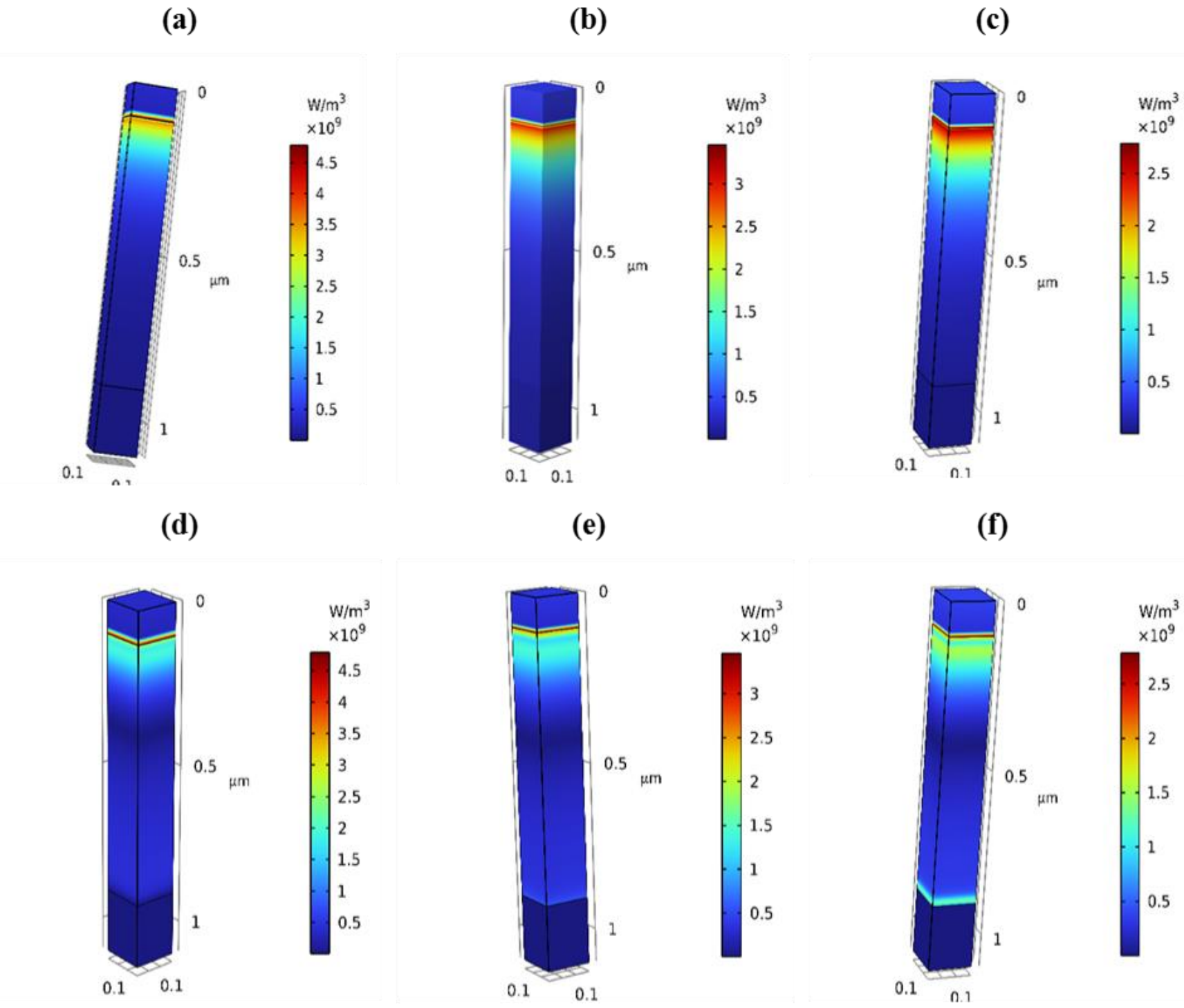

**Fig. 9:** Three-dimensional nonradiative recombination heating maps (W/m$^3$) for $MgXS_3$ (X = Ti, Zr, Hf). At V = 0 V: (a) $MgTiS_3$, (b) $MgZrS_3$, (c) $MgHfS_3$. At the optimum operating voltage, Vopt: (d) $MgTiS_3$, (e) $MgZrS_3$, (f) $MgHfS_3$.

In the analysis from Fig. 9(d) to 9(f), at the optimum operating voltage (Vopt), the heating is not restricted to a narrow interfacial region. A larger fraction of the absorber exhibits moderate heat generation, while the maximum intensity remains close to the front interface, with peak values of about $4.77 \times 10^9$ W/m$^3$ for $MgTiS_3$, $3.45 \times 10^9$ W/m$^3$ for $MgZrS_3$, and $2.76 \times 10^9$ W/m$^3$ for $MgHfS_3$ devices. This behavior indicates a higher steady-state carrier population under operating bias, which increases electron–hole overlap and strengthens SRH recombination throughout regions where bulk traps are present [69]. Notably, $MgHfS_3$ also shows a pronounced heat-generation feature near the rear interface, reaching approximately $1.36 \times 10^9$ W/m$^3$. This suggests enhanced interfacial recombination near the back contact under forward operation, which can arise from carrier accumulation caused by imperfect contact selectivity or unfavorable band alignment that impedes extraction and increases local electron–hole coexistence at the interface [70].

### *3.7.2 Spatial distribution of Joule heating in the proposed devices*

Fig. 10 illustrates the three-dimensional distribution of Joule heating in $MgXS_3$(X = Ti, Zr, Hf) based devices under short-circuit operation (V = 0 V) and at the optimum operating voltage (Vopt). Joule heating originates from resistive power dissipation during carrier transport and therefore scales with current flow and local resistivity. As a result, it is generally more pronounced under short-circuit conditions, where the photocurrent is highest, and it becomes much weaker near operating points where current density is reduced [44,46].

From Fig. 10(a) to 10(c), the Joule heat source at V = 0 V is highly non-uniform and is concentrated at the contacts and at the front and rear interfaces, while the absorber bulk remains comparatively weakly heated. This interface and contact localization is physically reasonable because current crowding, transport barriers, and local electric-field enhancement are typically strongest near junction and electrode regions, which promotes higher J·E dissipation than in the absorber interior [44,71]. Quantitatively, for $MgTiS_3$ cell the extracted Joule-heating peaks are $5.15 \times 10^9$ W/m$^3$ at the front contact and $2.90 \times 10^9$ W/m$^3$ at the back contact, with higher interfacial values of $8.48 \times 10^9$ W/m$^3$ at the ZnSe/$MgTiS_3$ interface and $7.08 \times 10^9$ W/m$^3$ at the $MgTiS_3$/$Sb_2S_3$ interface. For $MgZrS_3$, the peaks are $3.16 \times 10^9$ W/m$^3$ (front contact) and $1.89 \times 10^9$ W/m$^3$ (back contact), while the strongest hotspot appears at the front interface with $1.05 \times 10^{10}$ W/m$^3$ and a smaller rear-interface contribution of $1.70 \times 10^9$ W/m$^3$. For $MgHfS_3$, the corresponding values are $2.75 \times 10^9$ W/m$^3$ (front contact) and $3.40 \times 10^8$ W/m$^3$ (back contact), with interfacial peaks of $5.98 \times 10^9$ W/m$^3$ (front interface) and $2.15 \times 10^9$ W/m$^3$ (rear interface). Overall, the short-circuit maps indicate that Joule losses are governed primarily by resistive dissipation near carrier injection and extraction regions rather than by the absorber bulk.

From Fig. 10(d) to 10(f), the Joule-heating magnitude decreases substantially at Vopt for all three absorbers, confirming that resistive dissipation is reduced when the operating current density is lower than at short circuit. For $MgTiS_3$ cell, the peaks drop to $1.08 \times 10^9$ W/m$^3$ (front contact), $6.03 \times 10^8$ W/m$^3$ (back contact), $7.10 \times 10^8$ W/m$^3$ (front interface), and $1.04 \times 10^9$ W/m$^3$ (rear interface). For $MgZrS_3$, the values reduce to $8.36 \times 10^8$ W/m$^3$ (front contact), $4.99 \times 10^8$ W/m$^3$ (back contact), $1.10 \times 10^9$ W/m$^3$ (front interface), and $1.02 \times 10^9$ W/m$^3$ (rear interface). For $MgHfS_3$, the corresponding peaks are $5.28 \times 10^8$ W/m$^3$ (front contact), $3.15 \times 10^8$ W/m$^3$ (back contact), $4.55 \times 10^8$ W/m$^3$ (front interface), and $4.56 \times 10^8$ W/m$^3$ (rear interface). Even though the overall Joule-heating level is much lower at Vopt, the remaining dissipation is still concentrated near the contacts and interfaces, since these regions typically exhibit the largest local electric fields and transport resistance, leading to localized J·E power loss even when the net current is reduced [44,46].

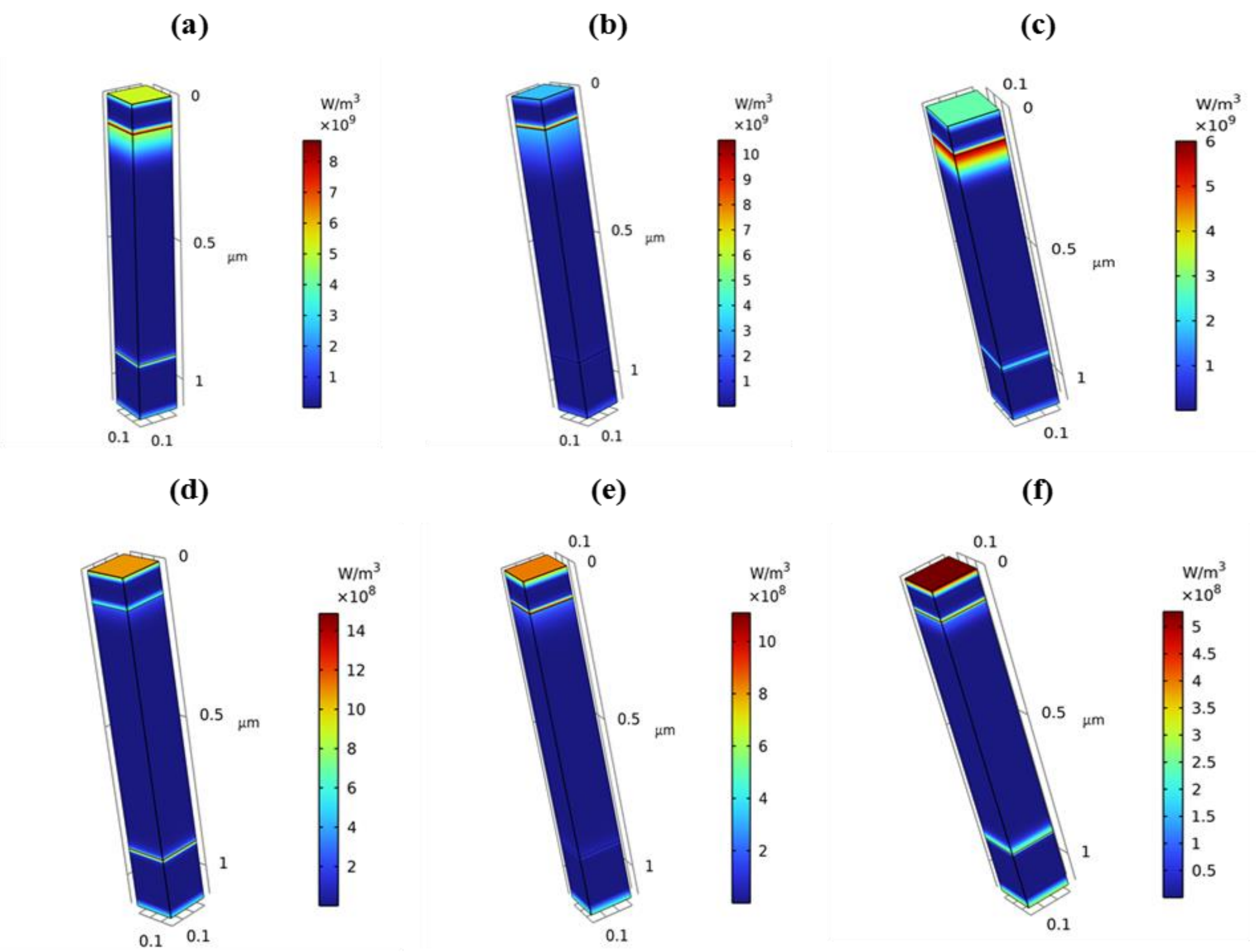

**Fig. 10:** Three-dimensional Joule heating maps for $MgXS_3$(X = Ti, Zr, Hf) devices: At V = 0 V fro: (a) $MgTiS_3$, (b) $MgZrS_3$, (c) $MgHfS_3$; and at the optimum operating voltage, Vopt for: (d) $MgTiS_3$, (e) $MgZrS_3$, (f) $MgHfS_3$.

### 3.7.3 Steady-state temperature-gradient and hot-spot temperature analysis

The thermal response of the proposed n-ZnSe/p-$MgXS_3$/p$^+$-$Sb_2S_3$ solar cells has been further examined by extracting the steady-state temperature-gradient magnitude, $|\nabla T|$, under 1-sun illumination using the coupled Semiconductor and Heat Transfer in Solids modules. As shown in Fig. 11, the steady-state temperature-gradient distributions of the $MgXS_3$(X = Ti, Zr, Hf)-based devices remain relatively low in the ZnSe window layer and $Sb_2S_3$ BSF region, indicating limited heat accumulation in these regions. In contrast, a comparatively stronger thermal response is observed within the absorber bulk, where non-radiative SRH recombination is expected to contribute significantly to internal heat generation [44]. However, a localized increases in the temperature-gradient magnitude occurs near the hetero-interfaces, which may arise from differences in thermal transport properties between adjacent layers, together with interface-related defects and band discontinuities [44, 72].

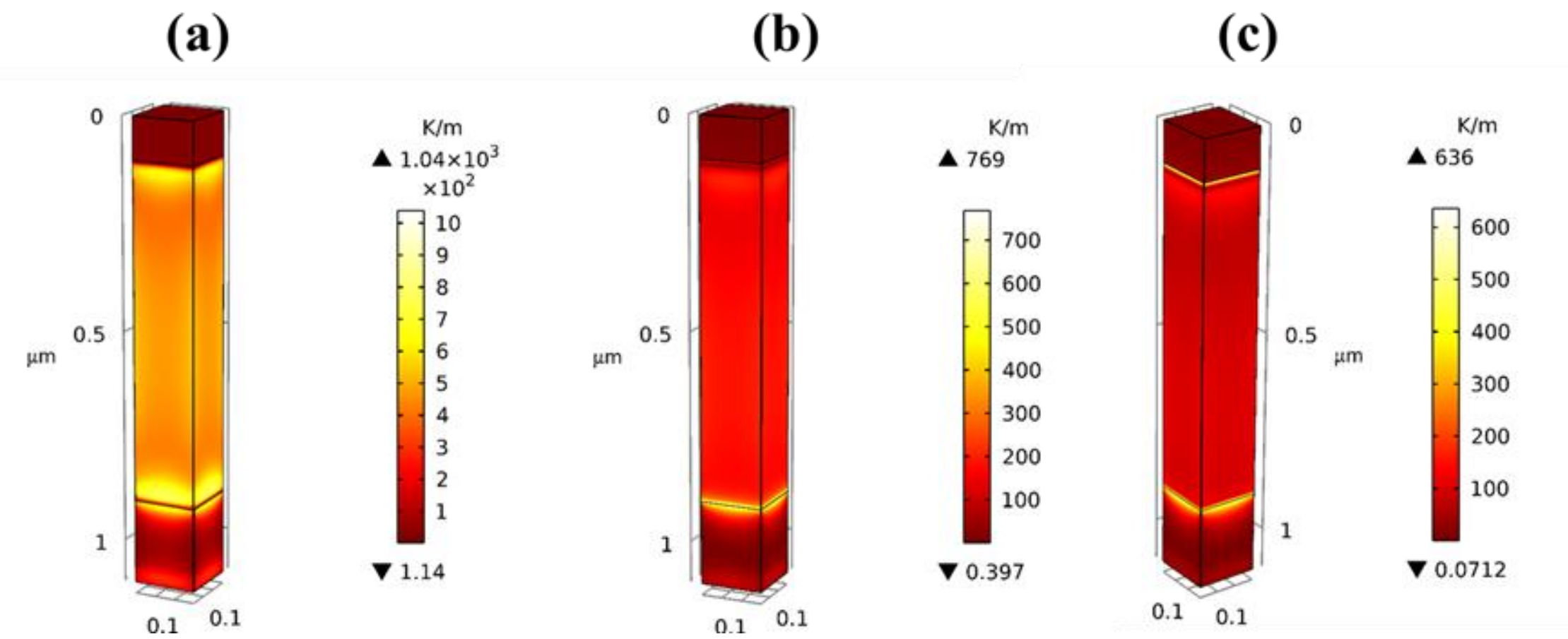

**Fig. 11**: Steady-state temperature-gradient magnitude distributions under 1-sun illumination for $ZnSe/MgXS_3/Sb_2S_3$ solar cells with (a) $MgTiS_3$, (b) $MgZrS_3$ and (c) $MgHfS_3$ absorber layers.

Taking 293.15 K as the reference temperature, the corresponding hot-spot temperatures in the simulated device with thickness of 1.1 µm are approximately 293.1512 K, 293.1509 K, and 293.1508 K for $MgTiS_3$, $MgZrS_3$ and $MgHfS_3$ cells. The estimated steady-state temperature increments are approximately 1.24 mK, 0.89 mK, and 0.75 mK, respectively of the corresponding devices. The $MgTiS_3$-based device exhibits the highest temperature-gradient magnitude reaching approximately $1.04\times10^3$ K $m^{-1}$. In comparison, the $MgZrS_3$ and $MgHfS_3$-based devices show lower maximum temperature-gradient values of about 769 and 636 K $m^{-1}$, respectively. This behavior of $MgTiS_3$-based device can be attributed to its higher photocurrent density, which may increase carrier-transport-related resistive heat dissipation and local heat-flow variation within the device [63].

## 4. Conclusion

A 3D simulation of $MgXS_3$(X = Hf, Zr, Ti) chalcogenide perovskite solar cells has been performed using COMSOL Multiphysics to assess their optical response, electrical output, and thermal loss behavior. All three device variants are evaluated using an identical device architecture in which a ZnSe window and an $Sb_2S_3$ BSF layers are maintained. By optimizing absorber thickness, doping concentration, and defect density in the three-dimensional device architecture, while keeping the window and BSF layers fixed, the devices achieve improved performance with peak PCE values of 28.16% for $MgHfS_3$, 28.18% for $MgZrS_3$, and 26.72% for $MgTiS_3$ devices. The QE analysis confirms the expected trade-off between spectral coverage and bandgap-limited voltage. $MgTiS_3$ device shows the strongest long-wavelength response, while $MgZrS_3$ and $MgHfS_3$ provide higher photovoltage and higher overall efficiency. Adding the $Sb_2S_3$ BSF also improves rear-side carrier collection, increasing the QE at representative wavelengths from 84.21% to 97.69% at 900 nm for $MgTiS_3$, from 83.48% to 96.56% at 800 nm for $MgZrS_3$, and from 83.01% to 98.02% at 700 nm for $MgHfS_3$ devices. Parasitic-resistance analysis displays that higher series resistance mainly lowers FF and power output, with little effect on $V_{OC}$ at open circuit, while the impact of shunt resistance becomes negligible once it exceeds approximately $10^4$ $\Omega$.cm$^2$ because leakage currents are then small compared with the diode and photocurrent contributions. Finally, the thermal analysis indicates that nonradiative recombination and resistive dissipation contribute to internal heat generation with localized hotspots concentrated near interfaces, highlighting regions that may be susceptible to heat-assisted degradation. Although $MgTiS_3$, $MgZrS_3$, and $MgHfS_3$ perovskites remain largely unexplored experimentally, the present study provides a predictive device-level assessment of their photovoltaic potential. The results highlight the promise of $MgXS_3$ chalcogenide perovskites as lead-free absorber materials and demonstrate that the proposed 3D opto-electro-thermal modeling framework can serve as an effective tool for guiding future experimental validation and device optimization.

## Conflicts of Interest

The authors declare no conflicts of interest.

*Corresponding author: E-mail: jak_apee@ru.ac.bd (Jaker Hossain)

## Data Availability Statement

The data that support the findings of this study are available from the corresponding author upon reasonable request.

## Declaration of generative AI and AI-assisted technologies

The authors declare that no AI or AI-assisted tools were used in preparing this manuscript.